\documentclass[english,twocolumn]{article}
\usepackage[utf8]{inputenc}
\usepackage[english]{babel}
\usepackage[big,online]{dgruyter}
\usepackage[sort&compress,square,numbers]{natbib}
\usepackage{bm}
\usepackage{amsmath,amsfonts,amssymb}
\usepackage{txfonts}
\usepackage{tabularx}
\usepackage{textcomp}
\usepackage{microtype}
\usepackage[separate-uncertainty=true]{siunitx} 

\begin{document}

 \journalname{Current~Directions~in~Biomedical~Engineering}
 \journalyear{2021}
 \journalvolume{4}
 \journalissue{1}
 \startpage{1}
 \openaccess
 \contributioncopyright[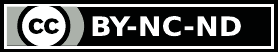]{2021}{The Author(s), published by De Gruyter.}{This work is licensed under the Creative Commons Attribution 4.0 License.}

  \title{Multi-Scale Input Strategies for Medulloblastoma Tumor Classification using Deep Transfer Learning}
  \runningtitle{Multi-Scale Input Strategies for Medulloblastoma Tumor Classification using Deep Transfer Learning}
  \subtitle{...}

  \author[1]{M. Bengs}
  \author[1]{S. Pant}
  \author[2,3,5]{M. Bockmayr}
  \author[2,3,4]{U. Schüller}
  \author[1]{A. Schlaefer}

  \runningauthor{M. Bengs et al.}

  \affil[1]{Institute of Medical Technology and Intelligent Systems, Hamburg University of Technology, Hamburg, Germany, \newline E-mail: marcel.bengs@tuhh.de }
  \affil[2]{Department of Pediatric Hematology and Oncology, University Medical Center Hamburg-Eppendorf, Martinistraße 52, Hamburg 20246, Germany  }
  \affil[3]{Research Institute Children's Cancer Center Hamburg, Martinistraße 52, Hamburg 20251, Germany }
  \affil[4]{Institute of Neuropathology, University Medical Center Hamburg-Eppendorf, Martinistraße 52, Hamburg 20246, Germany}
  \affil[5]{Mildred Scheel Cancer Career Center HaTriCS4, University Medical Center Hamburg-Eppendorf, 20246 Hamburg, Germany}

 \abstract{Medulloblastoma (MB) is a primary central nervous system tumor and the most common malignant brain cancer among children. Neuropathologists perform microscopic inspection of histopathological tissue slides under a microscope to assess the severity of the tumor. This is a time-consuming task and often infused with observer variability. Recently, pre-trained convolutional neural networks (CNN) have shown promising results for MB subtype classification. Typically, high-resolution images are divided into smaller tiles for classification, while the size of the tiles has not been systematically evaluated. We study the impact of tile size and input strategy and classify the two major histopathological subtypes—Classic and Demoplastic/Nodular. To this end, we use recently proposed EfficientNets and evaluate tiles with increasing size combined with various downsampling scales. Our results demonstrate using large input tiles pixels followed by intermediate downsampling and patch cropping significantly improves MB classification performance. Our top-performing method achieves the AUC-ROC value of 90.90\% compared to 84.53\% using the previous approach with smaller input tiles.}


\maketitle

\section{Introduction}
Medulloblastoma (MB) is the most common malignant brain tumor in children and a major cause of morbidity, as well as mortality in pediatric oncology  \cite{Pollack_Jakacki_2011}.
MBs are all classified as Grade IV tumors by the World Health Organization (WHO) \cite{louis2016}, indicating they are invasive and fast-growing. The 2016 edition of the World Health Organization Classification of Tumours of the Central Nervous System (CNS) has defined four histological subtypes of MB \cite{louis2016, northcott}— classic type (CMB), desmoplastic/nodular type (DN), MB with extensive nodularity (MBEN), and large cell anaplastic MB (LCA). Each subtype is associated with different prognoses and therapies, while early and precise diagnosis is vital for increasing the survival rates for patients \cite{northcott2019medulloblastoma}. \\
For establishing a diagnosis, a tissue specimen or biopsy sample is extracted from the suspected region of the brain. Then, neuropathologists assess the tissue slides under the microscope or digitize the magnified view to obtain an extremely high-resolution image which is also called Whole-Slide Image (WSI). 
To detect and discern different types and stages of tumors, they apply human-based decision rules based on their skills, experience, and knowledge to detect and discern different types and stages of tumors. However, the visual assessment of such tissue scans is a laborious and time-consuming task, which is also affected by inter-observer variability \cite{arevalo2014}. These problems have emphasized the requirement of automated decision support tool \cite{stenzinger2021artificial}. \\
One way to implement automated classification of MB subtypes is by means of manual feature extraction \cite{das2020}. Although this approach allows for promising results, manual feature extraction is task-dependent and requires strong domain expertise. In contrast to that,  Convolutional Neural Networks (CNNs) provide a more general approach and it has been demonstrated recently that CNNs outperform conventional methods in various pathological image analysis tasks \cite{alom2019,rachapudi2020improved}. While CNNs provide a general approach with superior performance in many learning tasks, they require a large number of training examples. However, in rare cancer such as MB, there are not enough training data available to train any powerful CNN architecture. To counter this problem, transfer learning is typically used in digital pathology \cite{phan2016transfer,kieffer2017convolutional}. \\
To mitigate the problem of scant training examples in MB classification, a recent study has compared various benchmark CNN architectures together with transfer learning \cite{bengs2021}. The study demonstrates that Efficinetnets \cite{tan2019efficientnet} with a larger input resolution outperform classical CNN architectures with smaller input resolutions. However, pre-trained CNNs are optimized for a fixed input resolution, e.g. $224\times224$ pixels \cite{tan2019efficientnet}, which conflicts with the high-resolution WSIs. Hence, WSIs are typically divided into several thousand tiles \cite{iizuka2020deep,alom2019}, which are processed with a deep learning approach afterwards. Here, the question arises which tile size to choose. 
\\
We systematically study the effect of tile size, image downsampling, and input strategy for the task of MB classification using pre-trained CNNs. We use a data set with WSI from 161 different patients and consider the task of differentiating between types CMB and DN.


\begin{figure*}
\centering
\includegraphics[width=0.80\textwidth]{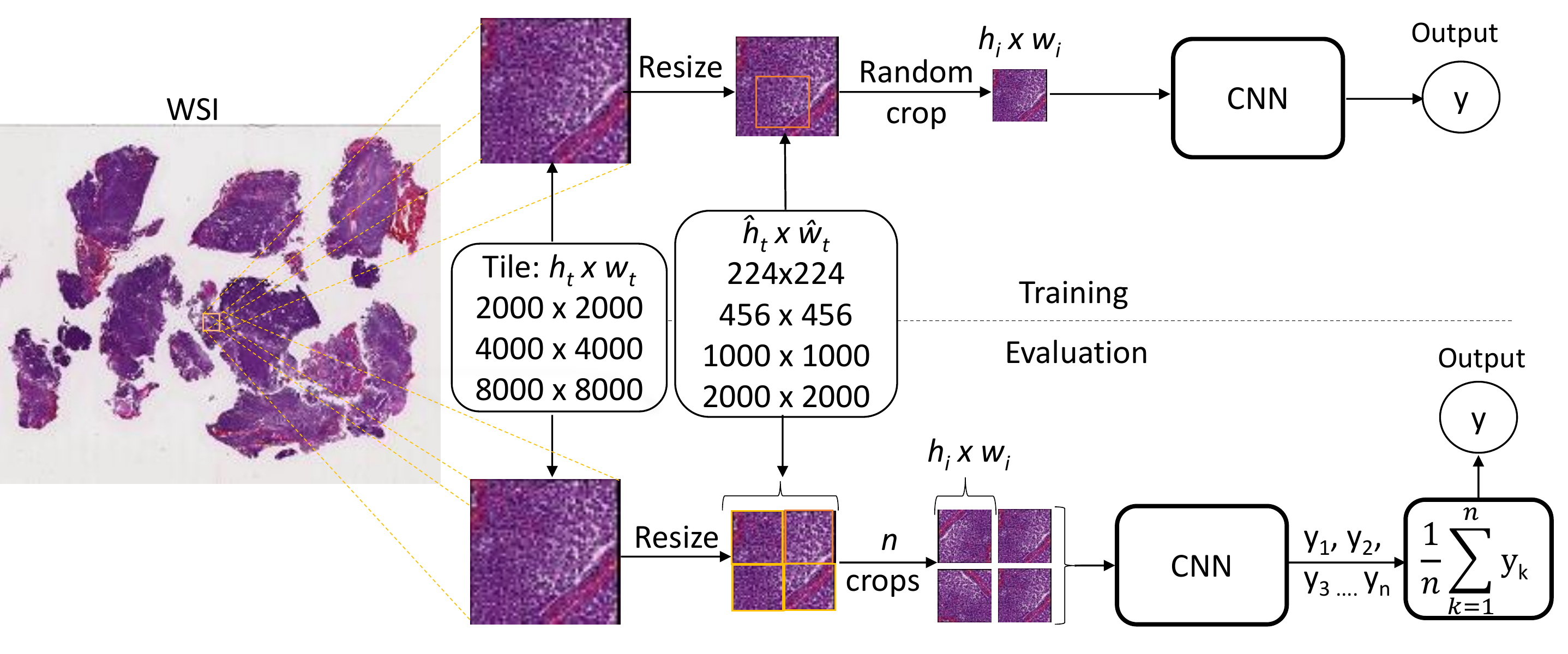}%
\caption{Our classification pipeline for the histological subtypes classic (CMB) and desmoplastic/nodular (DN). A tile ($h_{t} \times w_{t}$) is extracted from the WSI and is then downsampled to a size of $h_{t} \times w_{t}$. Afterwards, we take a randomly localized crop with the size of the network's input resolution $h_{i} \times w_{i}$ during training. For evaluation, we take $n$ ordered crops and average all crop prediction to obtain one final prediction $y$.}
\label{fig:pipeline}
\end{figure*}

\section{Materials and Methods} 
\subsection{Data Set} 
We use a dataset collected from 12 clinical sites in Germany from 1989-2011. All
local institutional guidelines were followed including informed consent from the
patients. Slides were stained by hematoxylin and eosin (H \& E) and
scanned at the same institution with a magnification of 200x. Neuropathologists have labeled the images as Classic (CMB) or Desmoplastic/Nodular (DN). The data contains WSIs of 161 patients of which 103 cases are CMB and 58 are DN. Each WSI has more than one cancerous region. To generate a data set consisting of image tiles, neuropathologists examined the WSIs and identified representative cancerous regions. Afterwards, we extracted tiles with a size of $2000\times2000$ pixels from the cancerous regions. Each patient contains multiple labeled tiles. There are 1574 tiles for 103 patients with CMB and 1195 tiles for 58 patients with DN cases. \\
We evaluate three different extracted tile sizes ($h_{t} \times w_{t}$). Given an extracted tile with a size of $2000\times2000$, we crop larger tiles with a size of 
$4000\times4000$ pixels and $8000\times8000$ pixels such that the manually extracted tile is centered. In this way, all sets share the same center area, while the large ones also include more overall context. We evaluate our models based on tile classification performance.

\subsection{Deep Learning Methods} 
We follow the concept of a previous study on MB classification \cite{bengs2021} and consider pre-trained EfficientNets \cite{tan2019efficientnet}. The key advantage of EfficientNets is the compound scaling method, which uniformly scales network width, depth, and input resolution starting with the baseline EfficientNet-B0. Considering the findings of previous works \cite{bengs2021}, we focus on EfficientNet-B0 (E\#Net-B0) with an input resolution of $224\times224$ and EfficientNet-B5 (E\#Net-B5) with an input resolution of $456\times456$. Note, we use architectures pre-trained on ImageNet. \\
Next, we study the relation between tile size, input strategy, and the corresponding classification performance. Our general classification pipeline is shown in Figure \ref{fig:pipeline}. \\
Given the sets with different tile sizes, we first follow the previous approach \cite{bengs2021} and simply downsample an entire image tile to the corresponding input resolution ($h_{i} \times w_{i}$) of a CNN. This leads to extreme downsampling of the larger tiles, especially for $8000\times8000$. Hence, we also consider an additional input strategy, where we first downsample the image tile to an intermediate resolution ($\hat{h}_{t}\times \hat{w}_{t}$) larger than the CNN input. Then, during training, we randomly crop input patches ($h_{i} \times w_{i}$) from the intermediate tiles, and during evaluation we take multiple ordered crops and average the predictions for an intermediate tile afterwards. We consider three different square intermediate tile sizes $\hat{h}_{t},\hat{w}_{t}\in\{456,1000,2000\}$. Note, when we combine the extracted tiles with a size of $2000\times2000$ with the intermediate tile size of $2000\times2000$ no downsampling is performed. \\
We randomly split our data based on patients and consider 10-fold cross-validation. In each fold, data is divided into a training set comprising 139 patients, and a test and validation set comprising of 7 patients each. The test and validation subsets all consist of five and two cases for type CMB and DN, respectively. To counter class inbalance during training, we weight the loss of the individual classes inversely proportional to samples of each class. We employ data augmentation during training, using brightness, contrast, saturation, and hue augmentation as well as random horizontal and vertical flipping of the images. The training is implemented with 300 epochs for all 10-folds with a batch size of 15. 

\section{Results}
We report report the area under the receiver operating curve (AUC) with $\SI{95}{\percent}$ confidence intervals (CI) using bias-corrected and accelerated bootstrapping with $n_{\mathit{CI}} = \num{10000}$ bootstrap samples in Table \ref{tab:All-networks-with metrics}. For testing of significance, we use a permutation test with $n_{\mathit{P}} = \num{10000}$ samples and a significance level of $\alpha = 5\%$ \cite{efron1994introduction}. Our results show that E\#Net-B5 outperforms E\#Net-B0 for all our experiments, except for an intermediate tile resolution of $465\times456$. Also, our results demonstrate that using a tile with size of $4000\times4000$ pixels, downsampled to an intermediate size of $2000\times2000$ pixels works best and significantly ($p<0.05$) outperforms the previous approach \cite{bengs2021} that used a smaller tile size of $2000\times2000$ px downsampled to $456\times456$ px. 

\begin{table}
\caption{Results for all experiments given in percent. The best performing method is shown in bold. 95 \% CIs are provided in brackets. Note, the input resolution of E\#Net-B0 and E\#Net-B5 are $224\times224$ px and $456 \times 456$ px, respectively. \label{tab:All-networks-with metrics}}
\begin{center}
\begin{tabular}{c c l l}
     &  &  E\#NET-B0  & E\#Net-B5  \\  
   Tile & $\hat{h}_{t}\times \hat{w}_{t}$ &  AUC & AUC \\     
\hline 

$2000$                      & $224\times224$ & $80.24(77-83)$\cite{bengs2021}  & $-$    \\
\multirow{2}{*}{$\times$} 	& $456\times456$ & $84.91(82-87)$ & $84.53(81-87)$\cite{bengs2021}       \\
							& $1000 \times 1000$ & $84.29(82-87)$ & $86.93(84-89)$      \\
 	$2000$					& $2000 \times 2000$ & $79.92(77-83)$ & $81.73(79-84)$   \\

 \hline
$4000   $                    & $224\times224$ & $82.67(81-86)$ & $-$   \\
\multirow{2}{*}{$\times$}	 & $456 \times 456$ & $82.97(79-85)$ & $85.27(82-88)$    \\
                             & $1000 \times 1000$ & $84.03(82-87)$ & $89.63(87-91)$    \\
 $4000$				         & $\mathbf{2000 \times 2000}$  & $86.42(84-88)$ & $\mathbf{90.90(89-93)}$     \\
	
 \hline
$8000 $                      & $224 \times 224$  & $77.72(74-80)$ & $-$  \\
 \multirow{2}{*}{ $\times$}	 & $456 \times 456$ & $82.95(79-85)$ & $83.74(81-86) $  \\
                             & $1000 \times 1000$ & $84.03(82-86)$ & $88.86(87-91)$  \\
 $8000$					     & $2000 \times 2000$ & $84.59(81-86) $ & $90.15(88-92)$   \\ 

\end{tabular}
\end{center}
\end{table}

\section{Discussion}
We consider MB subtype classification using pre-trained EfficientNets and study the impact of input patches with different scales and global context for this task. Our results in Table \ref{tab:All-networks-with metrics} demonstrate that using the previous approach \cite{bengs2021} with larger tiles downsampled to the network input resolution only lead to minor performance improvements for a tile size of $4000\times4000$ px. However, for extremely large tiles (8000 $\times$ 8000 px) performance is even reduced. This indicates hat using larger tiles is beneficial, however, when too much downsampling is performed relevant feature are lost. Similar, when no downsampling is performed performance is reduced, i.e., in the case of a tile size and an intermediate resolution of $2000\times 2000$ px. This indicates that here the global context is missing, while high-resolution information is preserved. This demands a method to preserve both global context without sacrificing the fine-grained details. Our results highlight that taking larger tiles, followed by intermediate downsampling and multi-cropping during training enables the right trade-off between the exploitation of global context and the preservation of detailed information. Our results demonstrate that this significantly improves the classification performance. Also, this superior performance might be linked to a simple version of multiple-instance learning (MIL) \cite{couture2018multiple}; the predictions are averaged in our study which acts as a pooling function in MIL terminologies. So far, we only consider MIL during evaluation, and considering more advanced versions of MIL during training like attention-based MIL \cite{ilse2018attention} could be lead to promising results. Also, WSI classification remains an open challenge and could be addressed by combining our findings with recent works on WSI classification \cite{iizuka2020deep,lu2021data}. 

\section{Conclusion}
We address the task of MB tumor classification and study the impact of input patches with different scales and global contexts. Our results highlight that including more overall image context is beneficial. However, simply downsampling larger tiles that cover enlarged image areas directly to the input resolution of a CNN does not lead to any performance improvement. Instead, downsampling to an intermediate resolution followed by a multi-cropping strategy significantly boosts performance. Future work could focus on evaluating more advanced MIL techniques and on classifying all subtypes of MB using a larger data set.

\textbf{Acknowledgments.} This work was partially supported by the Hamburg University of Technology i$^{3}$ initiative. 

\bibliographystyle{abbrvnat}
\bibliography{literature}   

\end{document}